\documentclass[twocolumn,epjc3]{svjour3}  

\smartqed  
\RequirePackage{graphicx}
\RequirePackage{amsmath}

\journalname{Eur. Phys. J. A}

\begin{document}

\title{Elastic Positron-Proton Scattering at Low Q$^2$}
\titlerunning{Elastic Positron-Proton Scattering}        

\author{Tyler~J.~Hague\thanksref{NCAT,e1}
        \and
        Dipangkar~Dutta\thanksref{MSU}
        \and
        Douglas~W.~Higinbotham\thanksref{jlab}
        \and
        Xinzhan~Bai\thanksref{UVA}
        \and
        Haiyan~Gao\thanksref{duke,triangle}
        \and
        Ashot~Gasparian\thanksref{NCAT}
        \and
        Kondo~Gnanvo\thanksref{UVA}
        \and
        Vladimir~Khachatryan\thanksref{duke,triangle}
        \and
        Mahbub~Khandaker\thanksref{ES}
        \and
        Nilanga~Liyanage\thanksref{UVA}
        \and
        Eugene~Pasyuk\thanksref{jlab}
        \and
        Chao~Peng\thanksref{ANL}
        \and
        Weizhi~Xiong\thanksref{SU}
        \and
        Jingyi~Zhou\thanksref{duke,triangle}
}

\thankstext{e1}{e-mail: tjhague@gmail.com}

\institute{North Carolina A\&T State University, Greensboro, NC 27411 \label{NCAT}
           \and
           Jefferson Lab, Newport News, VA 23601\label{jlab}
           \and
           Department of Physics and Astronomy, Mississippi State University, Starkville, MS 39762, USA\label{MSU}
           \and
           Department of Physics, Duke University, Durham, NC 27708, USA\label{duke}
           \and
           Triangle Universities Nuclear Laboratory, Durham, NC 27708, USA\label{triangle}
           \and
           Department of Physics, University of Virginia, Charlottessville, VA 22904, USA\label{UVA}
           \and
           Energy Systems, Davis, CA 95616, USA\label{ES}
           \and
           Physics Division, Argonne National Laboratory, Lemont, IL 60439, USA\label{ANL}
           \and
           Department of Physics, Syracuse University, Syracuse, NY 13244, USA\label{SU}
}

\date{Received: date / Accepted: date}

\maketitle

\begin{abstract}

Systematic differences in the the proton's charge radius, as determined by ordinary atoms and muonic atoms, have caused a resurgence of interest in elastic lepton scattering measurements.
The proton's charge radius, defined as the slope of the charge form factor at Q$^2$=0, does not depend on the probe.
Any difference in the apparent size of the proton, when determined from ordinary versus muonic hydrogen, could point to new physics or need for the higher order corrections. 
While recent measurements seem to now be in agreement, there is to date no high precision elastic scattering data with both electrons and positrons. 
A high precision proton radius measurement could be performed in Hall B at Jefferson Lab with a positron beam and the calorimeter based setup of the PRad experiment. 
This measurement could also be extended to deuterons where a similar discrepancy has been observed between the muonic and electronic determination of deuteron charge radius. 
A new, high precision measurement with positrons, when viewed alongside electron scattering measurements and the forthcoming MUSE muon scattering measurement, could help provide new insights into the origins of the proton radius
puzzle, and also provide new experimental constraints on radiative correction calculations.
\end{abstract}

\section{Introduction}
\label{intro}

Elastic lepton scattering at low four-momentum transfer can be used to determine the charge and magnetic radii of a nucleus. 
For the special case of a lepton scattering from a spin-1/2 nucleus, such as the proton or $^3$He,
the cross section for the scattering process can be written as
\begin{align}
    \sigma & =  \sigma_\mathrm{Mott} \times \nonumber \\
    & \left[\frac{G_E^2\left(Q^2\right)+\tau G_M^2\left(Q^2\right)}{1+\tau}+2\tau                           G_M^2\left(Q^2\right)\tan^2\left(\frac{\theta}{2}\right)\right] \text{,}
\label{eqrosen}
\end{align}
where $G_E(Q^2)$ and $G_M(Q^2)$, are the charge and magnetic form factors, $Q^2$ is the four momentum transfer squared, $m$ is the mass of the nucleus, and $\tau=\frac{Q^2}{4m^2}$.
By making measurements with multiple energies but over the same $Q^2$ ranges, the form factors $G_E(Q^2)$ and $G_M(Q^2)$ can be determined.
In the limit of $Q^2=0$, these form factors can be used to extract the charge radius, $r_E$, and magnetic radius, $r_M$, of the proton:
\begin{align}
 r_{E}^{p} \; \equiv \; \left( -6 \left.\frac{\mathrm{d} G_{E}^{p} (Q^2)} {\mathrm{d}Q^2} \right|_{Q^{2}=0} \right)^{1/2}, \nonumber\\
 \vspace{2em}
 r_{M}^{p} \; \equiv \; \left( \frac{-6}{\mu^{p}} \left.\frac{\mathrm{d} G_{M}^{p} (Q^2)} {\mathrm{d}Q^2} \right|_{Q^{2}=0} \right)^{1/2}
 \label{radii_extractions}
\end{align}
where $\mu^p$ is the magnetic moment of the proton. 
It is important to note, that this definition of the proton's radius is consistent with the definition used by the atomic and muonic lamb shift measurements~\cite{Miller:2018ybm}.
Experimentally, the data cannot extend to exactly $Q^2=0$, thus various methods of extrapolation are employed. 
To minimize the model dependence of these extrapolations, it is desirable for experiments to measure at $Q^2$ as low as achievable and over a sufficiently large $Q^2$ range.
As the measurement of the proton radius occurs at $Q^2=0$, achieving lower $Q^2$ minimizes the uncertainties with extrapolating the form factors to $Q^2=0$.
Measuring over a larger $Q^2$ range helps to minimize the uncertainty from the slope extraction when the form factors are fit.

In 2010, Lamb shift measurements in muonic hydrogen ($\mu$H)~\cite{Pohl:2010zza,Antognini:1900ns} with their unprecedented $<$0.1\% precision, reported a $r_E^p$ that was a combined eight-standard deviations smaller than the average value from all previous experiments. This discrepancy triggered the {\it ``proton radius puzzle''}~\cite{Carlson:2015jba,Miller:2018zfu,Gao:2021sml}.
The puzzle prompted new scattering experiments~\cite{Bernauer:2010wm,Bernauer:2013tpr,Mihovilovic:2019jiz} and numerous reanalyses of existing electron scattering data~\cite{Horbatsch:2015qda,Griffioen:2015hta,Higinbotham:2015rja,Lee:2015jqa,Graczyk:2014lba,Lorenz:2014vha,Horbatsch:2016ilr,Alarcon:2018zbz,Alarcon:2020kcz,Zhou:2018bon,Higinbotham:2019jzd,Mihovilovic:2020dmd,Borisyuk:2020pxo,Cui:2021vgm,Atac:2020hdq}.

The most recent electron scattering~\cite{Xiong:2019umf} and atomic hydrogen spectroscopy~\cite{Bezginov:2019mdi} results seem to be in agreement with the $\mu$H results~\cite{Hammer:2019uab}. 
Nonetheless, the new results do not rule out one of the original explanations for the proton radius puzzle~\cite{Carlson:2015jba}, a fundamental difference between electrons and muons that violates lepton universality. 
Previous experiments, performed in the '70s and '80s, showed that lepton universality holds at the 10\% level~\cite{Camilleri:1969ag,Braunstein:1972st}. 
While this is often theoretically accepted to be true in the standard model, there has yet to be any experimental tests of the precision to which this assumption is valid.
The MUSE experiment~\cite{Gilman:2013eiv,Gilman:2017hdr}, which has begun running at PSI, may be able to determine if universality holds, and thus if the proton radius puzzle is truly solved. 
However, it is highly desirable to verify the results from MUSE with high precision measurements with electrons and positrons.

The PRad experiment~\cite{Xiong:2019umf}, as will be discussed in detail in the next section, demonstrated the advantages of the calorimetric method in $e-p$ scattering experiments to measure $r_E^p$ with high accuracy. 
An upgraded experiment (PRad-II), which will reduce the overall experimental uncertainties by a factor of 3.8 compared to PRad has recently been proposed.
However, to maximize the impact, the experiment ideally would be performed with both electrons and positrons. 
Positrons can be used to measure the scattering cross section and subsequently extract $G_E^p$ and $r_E^p$. 
By combining the positron result with the PRad and PRad-II results, it would allow us to validate the radiative correction calculations for electron scattering that account for internal and external Bremsstrahlung suffered by the incident and scattered electrons and contributions from two-photon exchange (TPE) processes.
Studying the positron, PRad, and PRad-II measurement of $r_E^p$ alongside results from MUSE would allow for a sub-percent precision study of lepton universality.
Testing lepton universality in this way is especially interesting given the recent report of $3.1\sigma$ lepton universality breaking in beauty-quark decays by LHCb~\cite{Aaij:2021vac}.


\section{The PRad Experiment}

The original PRad experiment was designed to use a magnetic-spectrometer-free, calorimeter technique~\cite{Xiong:2019umf}. The innovative design of the PRad experiment enabled three major improvements over previous $e-p$ experiments:
(i)~The large angular acceptance ($0.7^{\circ} - 7.0^{\circ}$) of the hybrid calorimeter (HyCal) allowed for a $Q^2$~coverage spanning two orders of magnitude, $2.1~\times~10^{-4}$ to $6~\times~10^{-2}$~$(\mathrm{GeV/c})^2$, in the low $Q^2$ range. 
The single fixed location of \mbox{HyCal} eliminated the multitude of normalization parameters that have affected magnetic spectrometer based experiments, where the spectrometer must be physically moved to many different angles to cover the desired range in $Q^2$. 
In addition, the PRad experiment reached extreme forward scattering angles down to $0.7^{\circ}$ achieving the lowest $Q^2$~($2.1~\times~10^{-4}$ $(\mathrm{GeV/c})^2$~) in $e-p$ experiments, approximately 5 times lower than the previous lowest achieved $Q^2$~\cite{Mihovilovic:2016rkr}. 
Having a range of very low $Q^2$ data is critically important since $r^p_E$ is extracted by extrapolating to determine $\it G^{p}_{E}$($Q^2$) at $Q^2$~= 0. 
%
(ii)~The extracted $e-p$ ~cross sections were normalized to the well
known quantum electrodynamics process - $e^- e^- \to e^- e^- $~M{\o}ller~scattering from the atomic electrons~- which was measured simultaneously with the $e-p$ within the same detector acceptance.
This leads to a significant reduction in the systematic uncertainties of measuring the $e-p$ ~cross sections.
(iii)~The background generated from the target windows, one of the dominant sources of systematic uncertainty for all previous $e-p$ experiments, is highly suppressed in the PRad experiment.

\begin{figure*}[htb]
    \centering
    \includegraphics[width=0.75\linewidth]{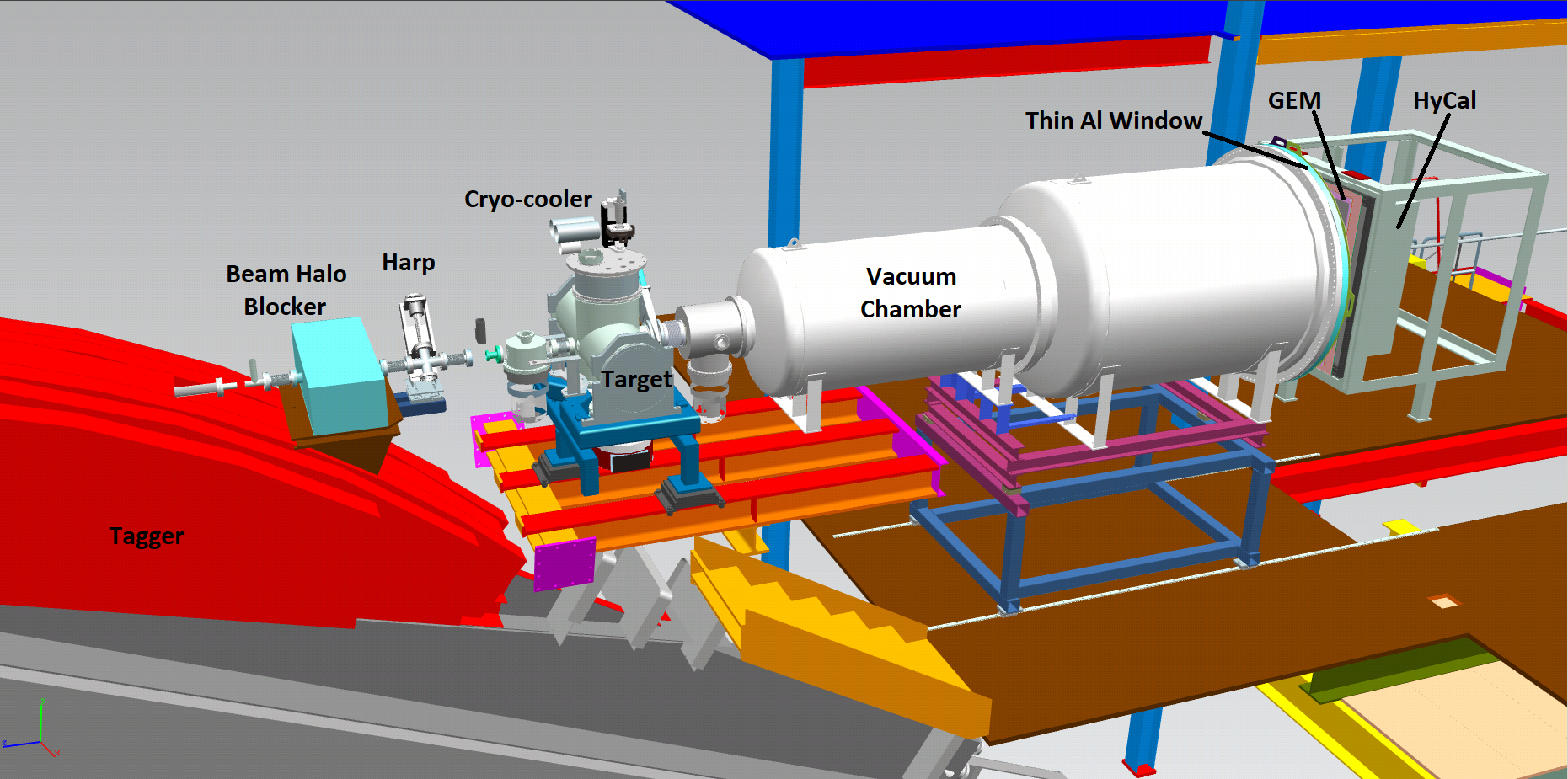}
\caption{A schematic layout of the PRad experimental setup in \mbox{Hall B} at Jefferson Lab, with the electron beam incident from the left. The key beam line elements are shown along with the windowless hydrogen gas target, the two-segment vacuum chamber, and the two detector systems, GEM and HyCal.} 
\label{fig:exp_setup}
\end{figure*}

\begin{figure*}[htb]
    \centering
    \includegraphics[width=\linewidth]{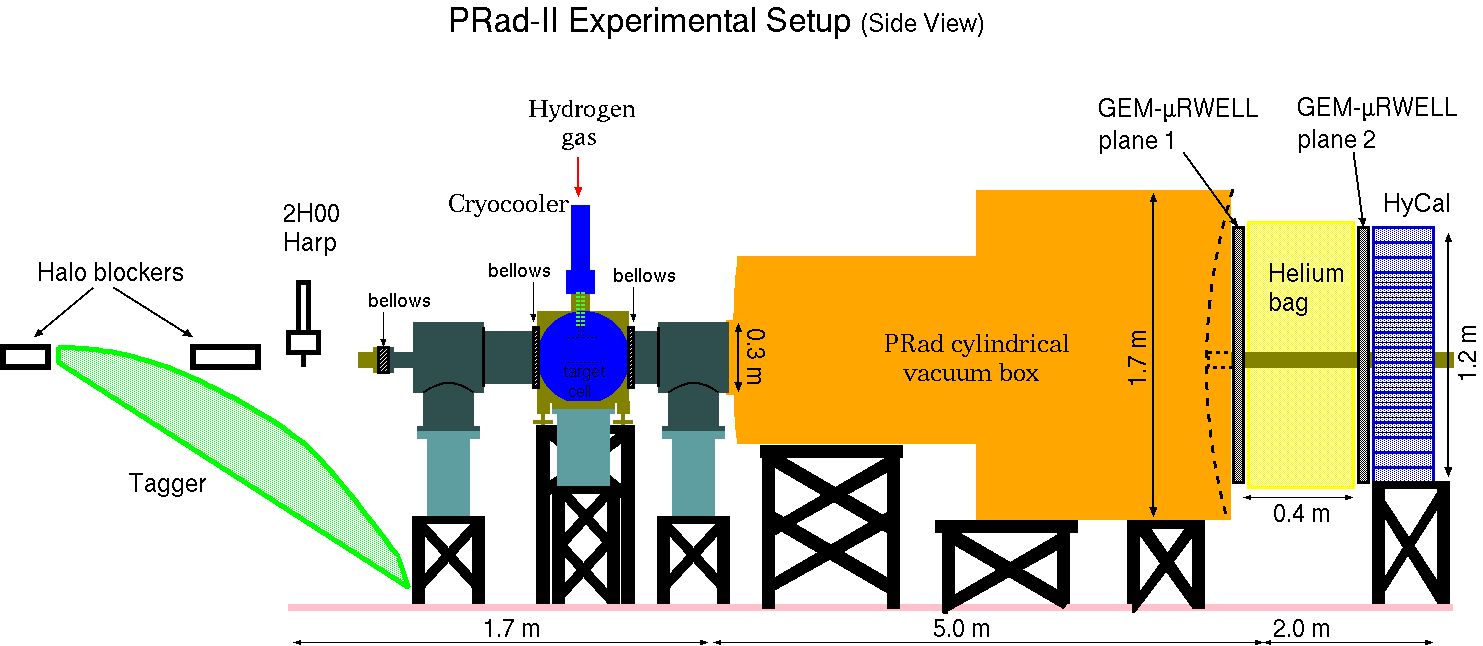}
    \caption{The proposed experimental setup for PRad-II.~\cite{Gasparian:2020hog}}
    \label{fig:PRad2_setup}
\end{figure*}

The PRad experimental apparatus consisted of the following five main elements (see Fig.~\ref{fig:exp_setup}): (i)~a 4-cm-long, windowless, cryo-cooled hydrogen (H$_2$) gas flow target with a density of \mbox{$2 \times 10^{18}$} 
atoms/cm$^2$~\cite{Pierce:2021vkh}. It eliminated the beam background from the target windows and was the first such target used in non storage-ring $e-p$ experiments;
(ii)~the high resolution, large acceptance \mbox{HyCal} electromagnetic calorimeter~\cite{Gasparian:2002px,Gasparian:2004xa}. The complete azimuthal coverage of \mbox{HyCal} for the forward scattering angles allowed simultaneous detection of the pair of electrons from $e-e$ ~scattering, for the first time in these types of measurements; 
(iii)~a single plane of coordinate detectors made of two high resolution $X-Y$ gas electron multipliers (GEM) located in front of \mbox{HyCal}; and
(iv)~a two-section vacuum chamber spanning the 5.5~m distance from the target to the detectors.  And finally (v) the low emittance beam provided by the CEBAF accelerator without which the experiment wouldn't be possible~\cite{Leemann:2001dg,Grames:2010zz}.

The PRad experiment was the first electron scattering experiment to utilize a new technique with completely different systematics compared to all previous magnetic-spectrometer-based $e-p$ experiments. 
The first generation PRad experiment was able to determine the proton radius to $\pm0.007_{\mathrm{stat}}\pm0.012_{\mathrm{syst}}$ fm~\cite{Xiong:2019umf,Yan:2018bez}. 
Thus, PRad demonstrated the validity and advantages of the new calorimetric technique, but further improvements are possible. 

 
\section{PRad-II and DRad}
 The second generation experiment -- PRad-II, which will reduce the overall experimental uncertainties by a factor of 3.8 compared to PRad, has been approved by the JLab 2020 Program Advisory Committee (PAC) with an A rating. PRad-II will cover the range of $Q^2$ range from $4\times10^{-5}$ GeV$^2$ to $3\times10^{-4}$ with only three settings, allowing a more accurate and robust extraction of the proton radius. This new experiments will push the precision of the proton radius extraction to 0.003~fm, allowing it to address possible systematic difference between $e-p$ and the $\mu$H experiments. 
 
 \begin{figure}[htb]
    \centering
    \includegraphics[width=\linewidth]{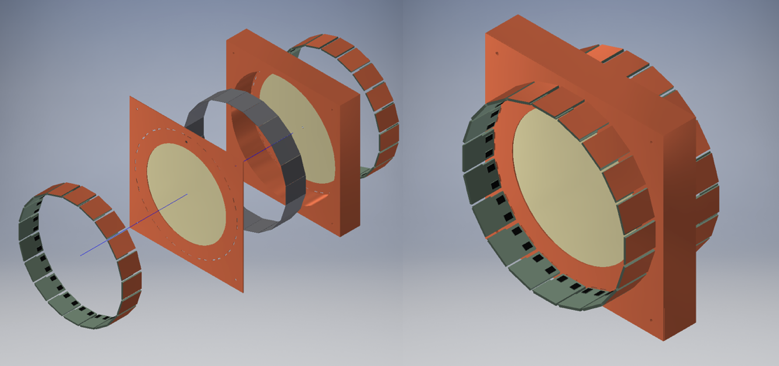}
    \caption{A schematic of the cylindrical recoil detector consisting of 20 silicon strip detector modules, held inside the target cell.}
    \label{fig:drad_recoil}
\end{figure}

Additionally, a proposal for a high precision elastic~$e-d$~scattering cross section measurement (DRad) at very low scattering angles 
$\theta_e$ from 0.7$^{\circ}$ to 6.0$^{\circ}$, corresponding to $Q^2$~=~$2~\times~10^{-4}$ to $5\times10^{-2}$ $(\mathrm{GeV/c})^2$, using the PRad-II experimental setup has also been submitted to the 2020 PAC.
This experiment has one major modification to the PRad-II setup. To ensure the elasticity of the $e-d$~scattering process a low energy Si-based cylindrical recoil detector will be included within the windowless gas flow target cell (see Fig.~\ref{fig:drad_recoil}). As in the PRad experiment, to control the systematic uncertainties associated with measuring the absolute $e-d$ cross section, a well known QED process, $e-e$ ~M{\o}ller~scattering, will be simultaneously measured in this experiment. The DRad experiment will provide a new measurement of the deuteron radius with a precision of 0.4\%.~\cite{Zhou:2020cdt}
 

\section{The Proposed Experiment}

Jefferson Lab, with a positron beam, would be ideal for performing a high precision follow-up experiment to MUSE and the PRad family of experiments. The setup used for the PRad-II experiment in Hall B could be reused to measure the cross sections and extract the proton radius, thereby verifying whether the proton radius is identical when measured with electrons and positrons.

Positrons, being the same mass as electrons, follow the same scattering kinematics as electron-proton scattering. Thus, the success of the PRad experiment and expected results of PRad-II and DRad serve as proof-of-concept for the proposed experiment.

The proposed experiment would use the same setup as the PRad-II and DRad experiments, as shown in Fig. \ref{fig:PRad2_setup}. By using this improved setup, a positron scattering measurement of the proton radius would be able to reach a minimum $Q^2$ in the range of 10$^{-5}$ GeV$^2$. This setup also provides ready-made solutions to several systematic issues that existed in the original PRad experiment.

The PRad-II setup improves on the PRad setup by the use of two novel spacerless GEM detectors and a larger angle scintillator detector. This setup pushes HyCal back by $40~\mathrm{cm}$ to make room for the second GEM detector. The spacerless GEM technology will greatly reduce the inefficiencies that particularly affected the very forward scattering angles. Using two GEM detectors will also allow them to use each other to determine their efficiencies and avoid using the lower position resolution of HyCal for calibration. A helium bag will span the $40~\mathrm{cm}$ distance between these detectors. This spacing will allow for accurate target-$z$ resolution (where the interaction occurred along the length of the target), which will help to mitigate beamline backgrounds.

\begin{figure}[htb]
\centering
\includegraphics[width=\linewidth]{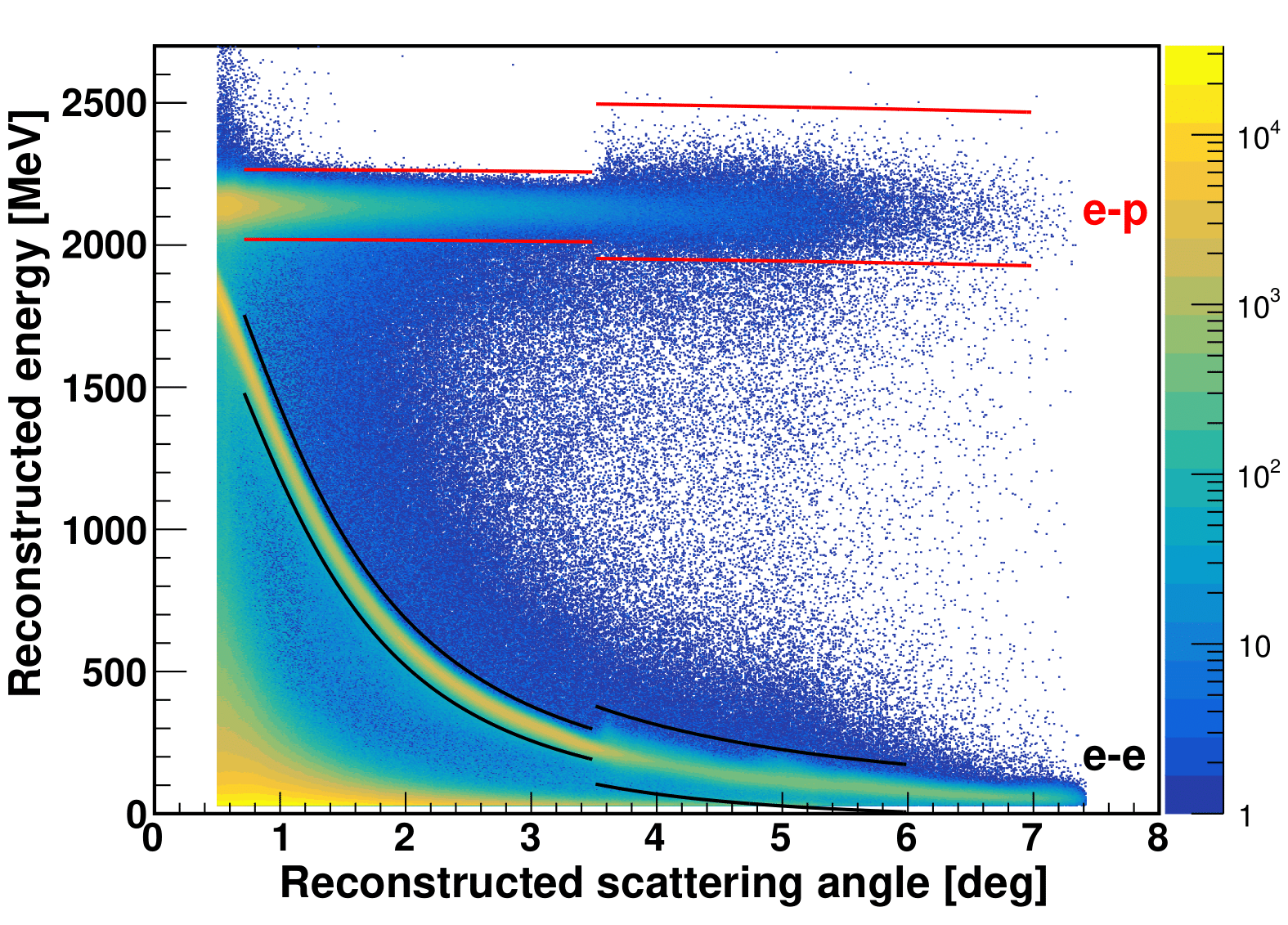}
\caption{The reconstructed energy vs angle for $e-p$ and $e-e$ events for the electron beam energy of 2.2 GeV. The red and black lines indicate the event selection for $e-p$ and $e-e$, respectively. The angles $\leq 3.5^{\circ}$ are covered by the PbWO$_4$ crystals and the rest by the Pb-glass part of \mbox{HyCal}.~\cite{Gasparian:2020hog}}
\label{PRad_AngleVE}
\end{figure}

As scattering angle decreases, the energy carried by electrons in both $e-p$ and $e-e$ scattering approach a value that is indistinguishable within the resolution of HyCal. This places a limit on the lowest usable scattering angles and, consequentially, lowest $Q^2$ measurable as shown in Fig. \ref{PRad_AngleVE}. A solution to this is to use double-arm ~M{\o}ller~scattering. At low scattering angles, however, the higher angle ~M{\o}ller~electron is outside of the acceptance of the main PRad setup. The PRad-II experiment will add a scintillator detector $25~\mathrm{cm}$ from the target in a cross shape. This addition will allow for the detection of the high scattering angle M{\o}ller electrons. Kinematic selection with these high scattering angle electrons will allow for clear discernment between $e-p$ and $e-e$ scattering.

The primary difference between the proposed experiment and PRad-II is the beam. 
A positron beam at JLab would be a secondary beam. 
This comes with the understanding that beam quality will not be as high as when using the main CEBAF electron beam. 
Two qualities of note are the anticipated beam emittance and flux. 
Higher beam emittance will increase the uncertainty surrounding the reconstruction of the vertex. 
The conversion from $e^-$ to $e^+$ will not be a highly efficient process, limiting the total flux of positrons that can be achieved.
The flux will also be limited by rejecting some positrons in order to constrain the emittance of the beam.
It will be possible to determine the emittance of the positron beam and should be possible increase to electron beam's emittance to match to minimize the systematic differences between the beams~\cite{Abbott:2016hyf,LDRD-positron,Accardi:2020swt}

\section{Luminosity Monitoring}

\begin{figure}
    \centering
        \includegraphics[width=0.827\linewidth]{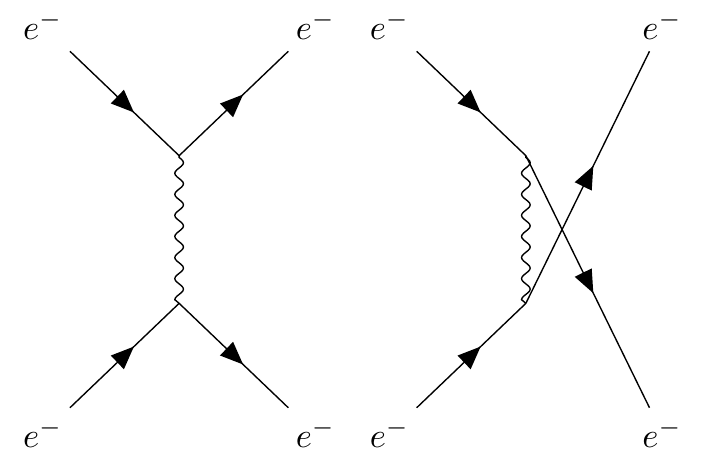} 
    \caption{Feynman diagrams of the leading order~M{\o}ller~scattering}
    \label{moller_feynman}
\end{figure}

\begin{figure}
    \centering
        \includegraphics[width=\linewidth]{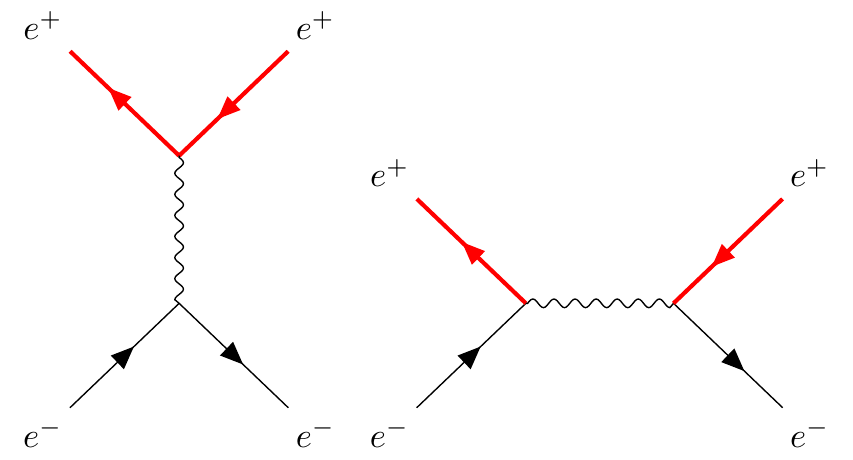}
    \caption{Feynman diagrams of the leading order Bhabha scattering}
    \label{bhabha_feynman}
\end{figure}

The PRad-type experiments have and will measure $e-e$ ~M{\o}ller~scattering (see Fig. \ref{moller_feynman}), a well known QED process, simultaneously with the $e-p$ cross section measurements. Being exactly calculable in the next-to-leading order makes the $e-e$ process an ideal candidate for monitoring the luminosity and controlling systematic uncertainties of the experiments. As an analog to this, the proposed positron scattering experiment would use $e^+e^- \to e^+e^-$ Bhabha scattering (see Fig. \ref{bhabha_feynman}) as a luminosity monitor.

The PRad-II detector setup is designed so that it could measure both $e^--p$ and $e^--e^-$ scattering simultaneously. 
The Bhabha scattering process follows identical kinematics to M{\o}ller scattering. 
Thus, a successful simultaneous measurement of $e^--p$ ($e^--d$) and $e^--e^-$ scattering in the PRad-II and DRad experiments will serve to prove the success of a simultaneous $e^+-p$ ($e^+-d$) and $e^+-e^-$ measurement.
By measuring an exactly calculable process at the same time as the elastic cross section, uncertainties associated with overall normalization are kept to a minimum.

\section{Analysis}

The analysis of this data is a two-step process: the $e^+-p$ elastic cross section extraction and the proton radius extraction. 
The cross section extraction requires a realistic simulation of both $e^+-p$ elastic scattering and of $e^+-e^-$ Bhabha scattering. 
The PRad experiment used a bin-by-bin M{\o}ller technique, which is analogous to a bin-by-bin Bhabha technique.
The benefit of the bin-by-bin technique is that detector inefficiencies will cancel in the analysis. 
However, a point-to-point $Q^2$-dependent uncertainty is introduced.
With the improved efficiencies of the spacerless GEM detectors, an integrated Bhabha technique could be used to normalize the $e^+-p$ elastic cross section. 
This involves using the Bhabha scattering measurement from the entire detector to normalize each $e^+-p$ cross section bin. 
The benefit of the integrated technique is that any $Q^2$-dependent uncertainties will be treated as overall normalization uncertainties, rather than point-to-point uncertainties.
The trade-off is that detector uncertainties will no longer cancel in the normalization.
With improved detector efficiency calculations, this can lead to an overall more precise extraction of the slope of $G_E^p$, and thus a more precise extraction of $r_p$.
By implementing the integrated Bhabha technique, the $e^+-p$ elastic cross section would be calculated as:
\begin{align}
    \left(\frac{\mathrm{d}\sigma}{\mathrm{d}\Omega}\right)^{\text{Born}}_{e^+p}&\left(\theta_i\right)=
    \frac{N_{e^+p}\left(\theta_i\right)\varepsilon^{\text{sim}}_{e^+p}}{N^{\text{sim}}_{e^+p}\left(\theta_i\right)\varepsilon_{e^+p}}\times \nonumber \\
    &\left[\frac{N^{\text{sim}}_{e^+e^-}\left(\text{PbO}_4\right)\varepsilon_{e^+e^-}}{N_{e^+e^-}\left(\text{PbO}_4\right)\varepsilon^{\text{sim}}_{e^+e^-}}\right]
    \left(\frac{\mathrm{d}\sigma}{\mathrm{d}\Omega}\right)^{\text{sim,Born}}_{e^+p}\left(\theta_i\right)\text{,}
    \label{XS_extract}
\end{align}
where $N$ are the counts of the respective processes from data and simulation without radiative corrections and $\varepsilon$ are the efficiencies and acceptances of the processes from data and simulation. The $e^+-e^-$ associated values are integrated over the PbWO$_4$ region of HyCal to form an overall normalization applied to all bins.

\begin{figure}[htb]
    \centering
    \includegraphics[width=\linewidth]{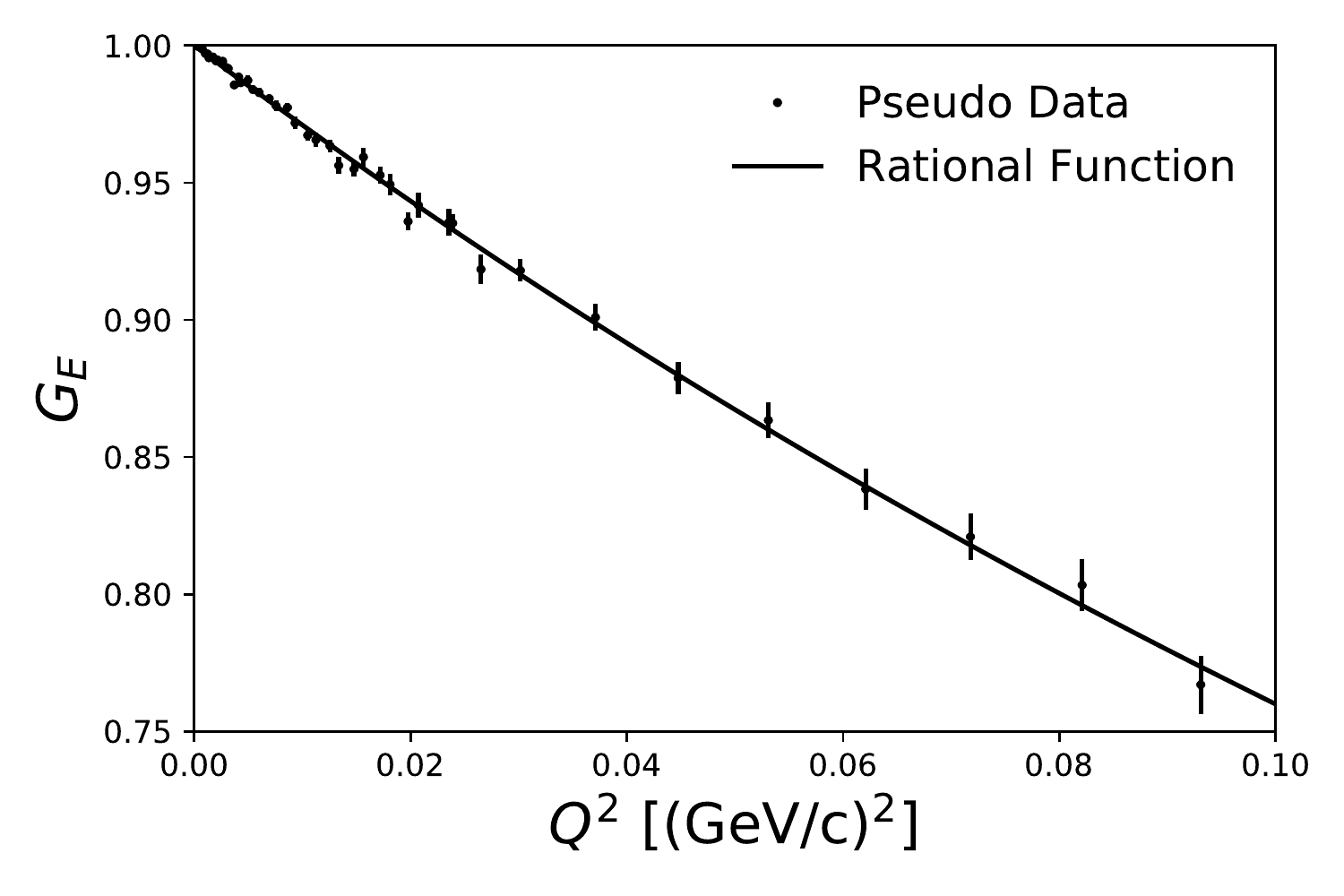}
    \caption{Shown are the expected precision of $e-p$ elastic scattering in Hall B using the PRad experimental setup.
    Data of this quality, would allow the proton radius to be extracted using a low order rational function and would achieve a precision approximately $\pm$ 0.007~fm statistical precision.    Using the proposal PRad-II setup, this precision can be improved by more than a factor of 3.8. }
    \label{low-q2-figure}
\end{figure}

\begin{figure}[htb]
\centering
\includegraphics[width=\linewidth]{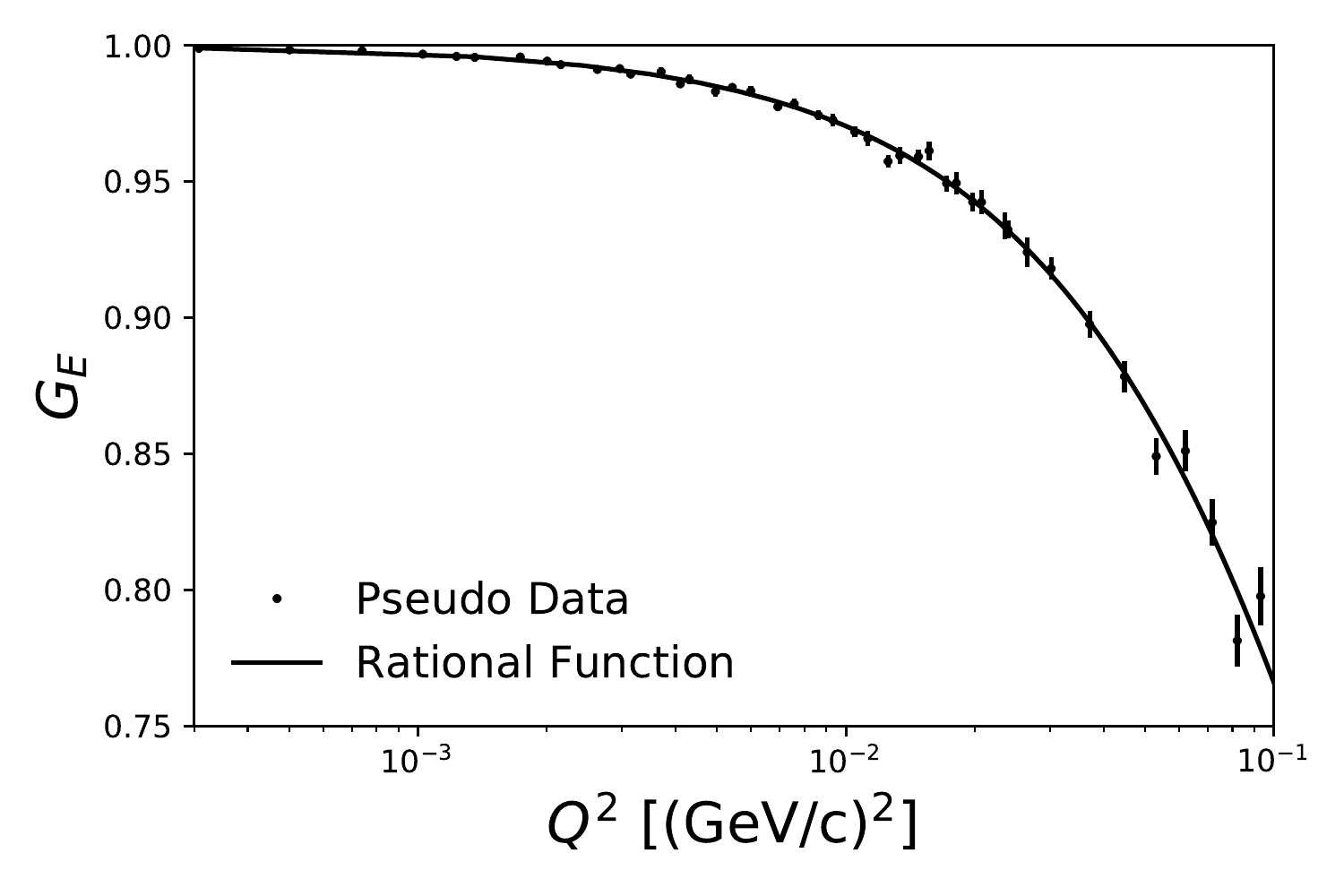}
\caption{Log scale version of Fig. \ref{low-q2-figure} to highlight the low $Q^2$ data.}
\label{low-q2-figure-log}
\end{figure}

After the cross section has been calculated, the electric form factor, $G_E^p$ is extracted from the Rosenbluth formula (Eq. \ref{eqrosen}) by assuming a model for $G_M^p$. 
This form factor must then be fit in order to determine the slope as $Q^2\to0$. 
As this is an extrapolation, great care must be taken to ensure that the functional form used yields an unbiased extraction of $r_E^p$. 
To do this, the procedure, as outlined in Ref.~\cite{Yan:2018bez}, needs to be followed. 
For the PRad experiment we found that the Rational $(1,1)$ function,
\begin{equation}
    f\left(Q^2\right)=nG_E\left(Q^2\right)=n\frac{1+p_1Q^2}{1+p_2Q^2}\text{,}
    \label{rational11}
\end{equation}
was the optimal choice and preliminary results indicate that this will again be the case for PRad-II. 
These studies are done by generating pseudo-data from various models of $G_E^p$ and then fitting the pseudo-data with several proposed functional forms. 
An example of this can be seen in Figs. \ref{low-q2-figure} and \ref{low-q2-figure-log}. The goodness of each fit was then determined by the Root Mean Square Error, $\text{RMSE}=\sqrt{\text{bias}^2+\sigma^2}$. Once the data has been fit, Eq. \ref{rational11} can be combined with Eq. \ref{radii_extractions} to yield a charge radius of
\begin{equation}
    r_E = \sqrt{6\left(p_2-p_1\right)}\text{.}
\end{equation}

The deuteron is a spin-1 nucleus, which means that an equation for the cross section different from Eq. \ref{eqrosen} is necessary. The cross section for $e-d$ elastic scattering is given by
\begin{equation}
    \sigma = \sigma_\mathrm{Mott} \times \left[A_d\left(Q^2\right)+B_d\left(Q^2\right)\tan^2\left(\frac{\theta}{2}\right)\right] \text{,}
\end{equation}
where $A_d$ and $B_d$ are structure functions that are defined in terms of three form factors (as opposed to two in the spin-1/2 case): the deuterons charge ($G_C^d$), magnetic dipole ($G_M^d$), and electric quadrupole ($G_Q^d$) form factors. The structure functions and form factors are related by
\begin{align}
    A_d\left(Q^2\right) = &\left(G_C^d\left(Q^2\right)\right)^2 + \frac{2}{3}\tau\left(G_M^d\left(Q^2\right)\right)^2 + \nonumber \\
    &\frac{8}{9}\tau^2\left(G_Q^d\left(Q^2\right)\right)^2 \text{,} \nonumber \\
    \vspace{2em}
    B_d\left(Q^2\right) = &\frac{4}{3}\tau\left(1+\tau\right)\left(G_M^d\left(Q^2\right)\right)^2 \text{.}
\end{align}
As in the proton case, this cross section would be normalized to Bhabha scattering that is measured simultaneously. The $e^+-d$ cross section would be extracted from the data using an analog to Eq.~\ref{XS_extract} where the $e^+-p$ terms are replaced by $e^+-d$ terms.

The deuteron charge radius is then given by
\begin{equation}
    r_{E}^{d} \; = \; \left( -6 \left.\frac{\mathrm{d} G_{C}^{d} (Q^2)} {\mathrm{d}Q^2} \right|_{Q^{2}=0} \right)^{1/2}\text{.}
\end{equation}
As this is calculated in the $Q^2\to0$ limit, like the proton radius, it is imperative to measure the cross section at as low $Q^2$ as possible. To extract $G_C^d$ from the measured cross section, data-driven models of $G_M^d$ and $G_Q^d$ would be used. The models that would be used can be explored in Appendix A of Ref.~\cite{Zhou:2020cdt}.

\section{Radiative Corrections}

A thorough analysis of radiative corrections is necessary to minimize systematic uncertainties on the final measurement. 
The PRad experiment has estimated the radiative correction related systematic uncertainty on the extracted $r_{E}^{p}$ to be $\delta r_{E}^{p} = 0.0069~{\rm fm}$. 
This estimation is anchored upon using analytic calculation results of the first-order radiative corrections from one-loop diagrams of $e-p$ and $e-e$ M{\o}ller scatterings. 
These are obtained within a covariant formalism and beyond the ultrarelativistic approximation~\cite{Akushevich:2015toa}, as well as using the method of Ref.~\cite{Arbuzov:2015vba} for evaluation of the contribution coming from higher order corrections. 
The estimated systematic uncertainties for both $e-p$ and $e-e$ scatterings are correlated and $Q^{2}$-dependent, where the $Q^{2}$-dependence is much larger for the corrections from the M{\o}ller process.

The PRad-II experiment will reduce the overall experimental uncertainties by a factor of 3.8 compared to PRad~\cite{Gasparian:2020hog}.
In order to succeed in this goal, it is necessary to perform radiative correction calculations to beyond the next-to-leading order and beyond the ultrarelativistic limit. 
This step is critical in order to sufficiently reduce the systematic uncertainty on $r_{E}^{p}$ associated with radiative corrections. 
Plans have already been put forward and detailed by PRad's theory colleagues to perform improved calculations at the next-to-next leading order level for elastic $e-p$ and M{\o}ller scatterings beyond ultrarelativistic limit for the PRad-II kinematics (including the contributions of the two-photon exchange processes).
This work will employ new calculation methods which they will develop in collaboration with a PSI-based group that is performing independent calculations. 
Similar radiative correction calculations will also be performed by PRad's theory colleagues for $e-d$ scattering pertaining to the DRad experiment.

Based on PRad's results, it is clear that the systematic uncertainties associated with radiative corrections stemming from $e^{+}-p$ $(e^{+}-d)$ and $e^{+}-e^{-}$ Bhabha scattering processes would also require thorough and meticulous studies. 
In this case, it is also critical to calculate them at and beyond the next-to-leading order and beyond the ultrarelativistic limit for these scattering processes. 
We believe that this goal may be accomplished after our theory colleagues successfully obtain the corresponding results for PRad-II.
The methods that are developed for calculations of the higher order radiative corrections in elastic $e-p$ and M{\o}ller scatterings should, in principle, be applicable to calculations for $e^{+}-p$ and Bhabha scatterings. 
Other approaches are also possible for achieving this goal, such as using the method developed in Ref.~\cite{Bucoveanu:2018soy} for numerical calculations of the second-order leptonic radiative corrections for lepton-proton scattering. 
We can additionally use the methods developed in Refs.~\cite{Becher:2007cu,Penin:2016wiw,Penin:2005kf,Bonciani:2006qu} for calculating the two-loop corrections to Bhabha scattering in the case that these methods are applicable to the kinematics of the PRad/PRad-II setup.

Of great interest is to use these results, in conjunction with the PRad-II results, to better understand two-photon exchange effects. A ratio of the results with standard radiative corrections applied, including soft two-photon exchange, will see the interference between one-photon and two-photon exchange cancel as these terms are charge odd. The ratio then will give direct access to hard two-photon exchange as any deviation from unity~\cite{Henderson:2016dea,Carlson:2007sp}.

\section{Impact}

The impact of this measurement is largely dependent on the findings of the PRad-II, DRad, and MUSE experiments. If the proton radius puzzle is not solved, then lepton universality will still be in question. This data would then provide a measure of the extent to which lepton measurements differ with electrons, positrons, and muons. 

On the other hand, it may be found that the proton radius puzzle is solved and that lepton universality still holds. If that is the case, then this data would be an ideal measure of positron radiative corrections, specifically internal and external Bremsstrahlung as well as two-photon exchange processes. Precise knowledge of the proton radius would allow for use of that as a fixed parameter for determining the correct radiative corrections to be applied.

\section{Summary}

Using the PRad setup in Hall B would allow for an extremely precise comparison of the proton and deuteron radii as extracted from positrons and electrons alongside world muon data.  While currently the initial proton radius puzzle seems to be solved, there is still a hint at a difference between muonic and atomic results which can only be resolved with higher precision experiments. In addition, even if the proton radius puzzle is solved, our understanding of radiative corrections and two-photon exchange processes can be improved by studying the differences between electrons and positrons.

\begin{acknowledgements}
This work is supported in part by the U.S. Department of Energy, Office of Science, Office of Nuclear Physics under contract DE-FG02-03ER41231 and DE-AC05-060R23177. This work is supported in part by NSF Grant NSF PHY-1812421.
\end{acknowledgements}

\bibliographystyle{spphys}       
\bibliography{bib2}   

\end{document}